\documentclass[%
 reprint,
showpacs,preprintnumbers,
 amsmath,amssymb,
 aps,
]{revtex4-1}

\usepackage{graphicx}
\usepackage{dcolumn}
\usepackage{bm}

\usepackage{subfigure}

\newcommand{\micron}{$\mathrm{\mu m}$}

\newcommand{\equref}[1]{Eq.~(\ref{#1})}
\newcommand{\figref}[1]{Fig.~\ref{#1}}
\newcommand{\Appref}[1]{Appendix.~\ref{#1}}

\newcommand{\abs}[1]{\left|{#1}\right|}

\begin{document}

\preprint{}

\title{Mesoscopic Turbulence and Local Order in Janus Particles\\ Self-Propelling under an AC Electric Field}

\author{Daiki Nishiguchi}
 \email{nishiguchi@daisy.phys.s.u-tokyo.ac.jp}
 \author{Masaki Sano}
 \affiliation{Department of Physics, The University of Tokyo, Hongo 7-3-1, Tokyo 113-0033, Japan}
 
\date{\today}

\begin{abstract}
To elucidate mechanisms of mesoscopic turbulence exhibited by active particles, we experimentally study turbulent states of non-living self-propelled particles. We realize an experimental system with dense suspensions of asymmetrical colloidal particles (Janus particles) self-propelling on a two-dimensional surface under an AC electric field. Velocity fields of the Janus particles in the crowded situation can be regarded as a sort of turbulence because it contains many vortices and their velocities change abruptly. Correlation functions of their velocity field reveal the coexistence of polar alignment and anti-parallel alignment interactions, which is considered to trigger mesoscopic turbulence. Probability distributions of local order parameters for polar and nematic orders indicate the formation of local clusters with particles moving in the same direction. A broad peak in the energy spectrum of the velocity field appears at the spatial scales where the polar alignment and the cluster formation are observed. Energy is injected at the particle scale and such conserved quantity as energy could be cascading toward the larger clusters.
 
\end{abstract}


\maketitle


\section{\label{intro}Introduction}
Self-propelled particles exist everywhere in nature. Their collective behaviors have been studied as subjects of nonequilibrium statistical physics \cite{Vicsek1995,Shimoyama1996}. The examples include swarming of bacteria \cite{Darnton2010,Zhang2010}, active turbulence of bacteria \cite{Wensink2012,Liu2012}, flocking of birds \cite{Cavagna2010}, and schooling of fish.  The swimming of such microswimmers as bacteria, algae, and spermatozoa at low Reynolds number has also attracted attention from  the viewpoint of hydrodynamics \cite{Purcell1976}.

Some studies have suggested power-law behavior in active turbulence, which is mesoscopic turbulence exhibited by active particles. Turbulent states of bacterial suspensions have been reported to show power laws in power spectra of their velocity fields \cite{Wensink2012,Liu2012}, which is analogous to usual fluid turbulence. Velocity fields of human crowds in a panic are also known to exhibit power laws in such quantities as structure function \cite{Helbing2007}. Power laws in power spectra are also reported in numerical studies on turbulent states of self-propelled rods model \cite{Wensink2012,Wensink2012b}, agent-based model \cite{Grossmann2014,Grossmann2015}, and slender rod-like microswimmers \cite{Saintillan2012}.

Conditions for the emergence of mesoscopic active turbulence have been investigated by numerical and theoretical studies. Wensink et al. \cite{Wensink2012,Wensink2012b} conducted numerical simulations on self-propelled rods interacting only through steric interactions and concluded that a range of aspect ratio of self-propelled rods that covers typical bacterial values can reproduce turbulent states similar to what was observed in their bacterial experiments, which implies that isotropic or spherical particles cannot exhibit turbulent states. However, in their model, hydrodynamic effects were dismissed although bacteria are swimming very close to each other in fluids. Gro\ss mann et al. \cite{Grossmann2014,Grossmann2015} simulated and constructed a kinetic theory on an agent-based model with short-range polar alignment and anti-parallel alignment at larger distances, which also reproduced turbulent states with power laws in the power spectra of their velocity fields. Although these works suggested sufficient conditions for active turbulence, they omit hydrodynamic viewpoint and thus it remains unclear whether such microscopic interactions as hydrodynamic interactions in microswimmer systems can induce such alignment effects leading to turbulent states. It is possible that even spherical microswimmers, whose aspect ratio is lower than the values for turbulent states reported in the simulations on self-propelled rods model \cite{Wensink2012,Wensink2012b}, can exhibit turbulent states by realizing appropriate alignment effects through any interactions other than steric ones. 

Experimental measurements on the power spectra of the velocity fields so far have been limited to systems of living bacteria, so there is a need to explore power spectra in non-living active systems. For the purpose of studying collective behaviors, various experimental systems with non-living self-propelled particles have been devised. These experiments can help us distinguish between phenomena originating from motility itself and those originating from biological activities. Shaken granular rods \cite{Narayan2007,Kumar2014} or shaken polar disks \cite{Deseigne2010} have been used to study collective behaviors, but it is difficult to prepare large-scale experiments to remove effects of boundaries. Experimental systems with micron-sized particles swimming in fluids have also been invented, although interactions of swimmers are more complicated than those in granular systems. Among such microswimmers, self-catalytic rods or colloids in hydrogen peroxide solution have been the most deeply investigated to demonstrate some fascinating characteristics of active Brownian particles systems \cite{Wang2006,Palacci2010,Ginot2015}, although it is difficult to realize steady states or vary the motilities of the particles instantaneously in these setups due to the finite chemical source of activity.

Considering these difficulties in other systems, experimental systems which are externally controllable by electric fields or lasers have been realized and studied \cite{Gangwal2008,Jiang2010a,Buttinoni2012,Bricard2013}. One example of such systems is that of Quincke rollers, which are symmetrical colloidal particles self-propelling under a DC electric field \cite{Liao2005,Jakli2008,Bricard2013}. When the particles in a hexadecane solution are placed under a DC electric field, the symmetry of the charge distribution is spontaneously broken, which causes the particles to rotate and move. It was found that they show transition from incoherent motion to macroscopically directed motion. Another example is a system of Janus particles, which are asymmetrical particles made of dielectric colloids whose hemispheres are covered with metal \cite{Gangwal2008}. When an AC electric field is applied to a suspension of the Janus particles, the difference in the dielectric constant of the two hemispheres induces asymmetrical electro-osmotic flow around them, which drives the Janus particles to move in a direction perpendicular to the electric field \cite{Squires2004,Squires2006,Gangwal2008}. When the suspension is sandwiched between two-dimensional electrodes as depicted in \figref{ExpSetup}, their motions are restricted in two-dimensional planes parallel to the electrodes \cite{Suzuki2011}.

In this paper, we study turbulent states of the Janus particles described above. Although experimental systems of the Janus particles has been realized, experiments with high number density and high voltage, which are necessary to observe collective dynamics with high motilities, have been challenging because of the technical difficulty in making the particles to swim smoothly without sticking to the electrodes. Hence previous studies focused on measuring properties of motions of single particle or doublets pivoting on an electrode \cite{Gangwal2008,Suzuki2011,Boymelgreen2014,Peng2014}. Here we improved the surface treatment of the Janus particles and the electrodes, which enables the particles to move at as fast as 61.4 $\mathrm{\mu m/s}$, which is approximately three times faster than the speed achieved by a previous study \cite{Gangwal2008} or typical bacteria such as \textit{Escherichia coli} and \textit{Bacillus subtilis}. We observed two-dimensional dynamics of the Janus particles at the projected area fraction of 0.24.

We obtained a power spectrum of the velocity field of dense suspension of self-propelling Janus particles, which has a broad peak. To elucidate the origin of the form of the power spectrum, we calculated velocity correlation functions and local order parameters for polar and nematic orders. The correlation functions reveal the coexistence of the polar alignment and the anti-parallel alignment interactions, which is analogous to those in the agent-based model by Gro\ss mann et al. \cite{Grossmann2014,Grossmann2015}. The order parameters indicate local cluster formation moving in the same direction. The range of wave numbers where the broad peak is observed corresponds to the spatial scales where the clusters are observed, which means that energy is injected and transferred by the particles themselves and the clusters in these scales.

\section{\label{Setup}Experimental Setup}
\begin{figure}[htbp]
 \begin{center}
    \includegraphics[width=\columnwidth]{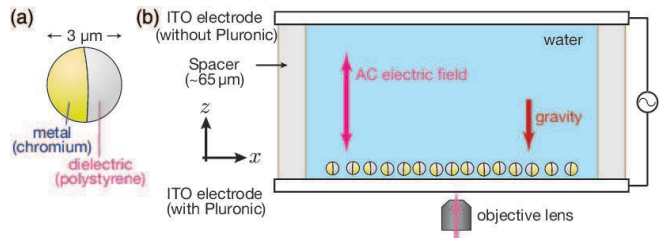}
 \end{center}
  \caption{\label{ExpSetup} (Color online) (a) Janus particles used in our experiments are polystyrene beads whose hemispheres are covered with chromium. (b) Suspension of the Janus particles were sandwiched between two horizontal transparent ITO electrodes. After sedimentation of the Janus particles, an AC electric field was applied in the $z$ direction and the Janus particles moved in quasi-two-dimensional horizontal plane ($xy$ plane). Images were captured through the bottom electrode using a microscope and a CCD camera.}
\end{figure}

Janus particles used in our experiments are spherical colloidal particles made of polystyrene whose hemispheres are covered with chromium (\figref{ExpSetup}(a)). To fabricate the Janus particles, purchased polystyrene colloidal particles (diameter 3 \micron, Polyscience, cat\#17134) were first placed on a glass slide to form a monolayer. Then chromium was deposited on this glass slide by thermal evaporation to form layers of thickness 20 nm on their hemispheres. The Janus particles on the glass slide were resuspended in deionized water and washed with sonication. During the washing procedure, water was once replaced with 5 \% solution of surfactant Pluronic F-127 (Invitrogen) to coat the surfaces of the colloidal particles. Excess surfactant was removed by replacing the medium after sonication and centrifugation.

In our experiments, as shown in \figref{ExpSetup}(b), suspension of Janus particles was sandwiched between two transparent electrodes (ITO glasses, Sanyo Shinku, Japan). The electrodes were separated by stretched Parafilm (thickness: $\sim65$ \micron). After waiting for a couple of minutes, all the Janus particles were sedimented to form a quasi-two-dimensional layer near the bottom electrode. When an AC electric field is applied, the Janus particles move around in this quasi-two-dimensional plane, which is perpendicular to the electric field. We observed their motions through the bottom electrode by using a microscope (Olympus, IX70) with a $\times 40$ objective lens (Olympus, LUCPLFLN, NA 0.60). Images were captured by a CCD camera (Imperx, IPX-VGA210-LMCN) at 200 Hz.

The origin of the self-propulsive forces of the Janus particles is the asymmetrical flow around the particles driven by induced charge electro-osmosis (ICEO). Because no net external forces are exerted on the whole system of the Janus particles and the fluid, total momentum of the particles and the surrounding fluid have to be conserved except in the vicinities of the electrodes. Therefore, the counter-action of these asymmetrical flows drives the Janus particles to move toward the polystyrene hemispheres \cite{Squires2004,Squires2006,Gangwal2008}. The motion of the Janus particles thus caused is called induced charge electrophoresis (ICEP). The boundary between the metal side and the polystyrene side of each Janus particle is always aligned to be parallel to the electric field, because of the hydrodynamic torque exerted by ICEO \cite{Kilic2011a}. Hence the self-propulsive force is always directed in the two-dimensional plane.

Surface of the bottom electrode was also coated with surfactant to avoid sticking. After washing the ITO glasses with sonication, they were treated with strong plasma cleaning. The cleaned ITO glasses were immersed in 5\% solution of the Pluronic F-127 for 4 hours and then the Pluronic solution was washed away by mildly pouring water onto the ITO glasses.

The gap of the two electrodes was approximately 65 \micron, and the amplitude of applied voltage was 15 V (30 V peak-to-peak) and its frequency was 1 kHz. Therefore the amplitude of the AC electric field was $2 \times 10^5$ V/m, which is one order of magnitude stronger than that in the previous studies \cite{Gangwal2008,Suzuki2011,Boymelgreen2014}. This led to higher motilities of the particles, because the self-propelling speed of the Janus particles is theoretically and experimentally proven to be proportional to the squared amplitude of the electric field \cite{Squires2006,Gangwal2008,Suzuki2011}. Another reason for choosing this large amplitude was to prevent aggregation in this high-density setup, which happened below 10 V at this frequency.

\begin{figure}[hbtp]
 \begin{center}
 \includegraphics[width=0.9\columnwidth]{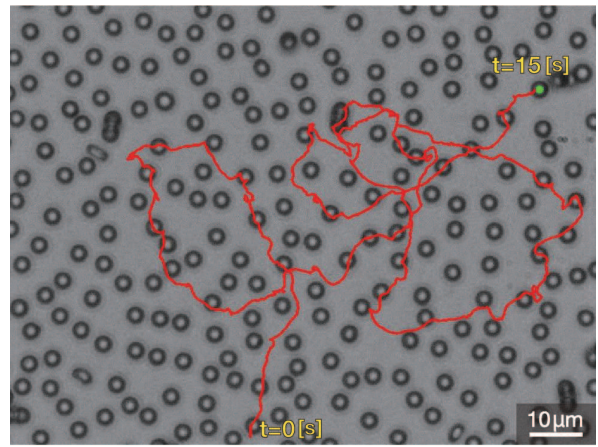}
 \end{center}
  \caption{\label{Trajectory} (Color online) A trajectory of a Janus particle for 15 seconds overlaid on the snapshot at the last moment. Due to rotational Brownian motion and frequent collisions of the particles, their trajectories are not straight but curved. The polarities of the particles are invisible with this magnification. The scale bar corresponds to 10 \micron. \cite{MovieTrajectory}}
\end{figure}

Captured movies were analyzed by detecting all the particles. After eliminating noise from the raw images, edges of the particles were detected and then binary images were obtained by filling holes of the edges. In order to separate the particles in collision, morphological operations were repeatedly applied to the binary images if necessary \cite{ImageProcessingMethod}. The magnification of $\times 40$ enabled high accuracy of detection, although the observation area was smaller compared with lower magnifications. All the particles were successfully tracked as long as they stay inside the frame (\figref{Trajectory}). However, the polarities of the particles are not detectable with this magnification due to the thin layers of chromium. The observation area was 120 $\mathrm{\mu m} \times 90 \mathrm{\mu m}$, which was centered at the sandwiched suspension with the diameter of approximately 5 mm and far away enough from the boundaries compared with the size of the particles to decrease the effects of boundaries. The spatial resolution of images was 0.19 $\mathrm{\mu m/pixel}$. Duration of the analyzed movie in this paper was 30.7 seconds (6094 frames). Because the self-propulsion speed was as high as 61.4 $\mathrm{\mu m/s}$ on average and the particles were densely packed, sufficient statistics were obtained from these data.

\section{\label{result}Results}
\subsection{Power spectrum of velocity field}
In the previous experimental studies on bacterial turbulence \cite{Wensink2012,Liu2012} and the numerical studies on a turbulent state of self-propelled rods \cite{Wensink2012,Wensink2012b}, it was suggested that power laws in the power spectra might be ubiquitous in the velocity fields of the turbulent states of self-propelled particles. Such power laws in power spectra are also found in fluid turbulence \cite{LandauFluidMech} or elastic turbulence \cite{Groisman2000}, although there are differences in the mechanisms and the exponents of these turbulent states. Power-law behavior has great significance because it might work as a clue for finding scale-free structures or conserved quantities. The velocity field of the Janus particles can be regarded as a sort of turbulence, because it contains various scales of vortices as we see in \Appref{AppendixVorticity}. Therefore, it is of considerable interest to examine how the power spectrum looks like in the Janus particles system.

\begin{figure}[hbtp]
 \begin{center}
  \includegraphics[width=\columnwidth]{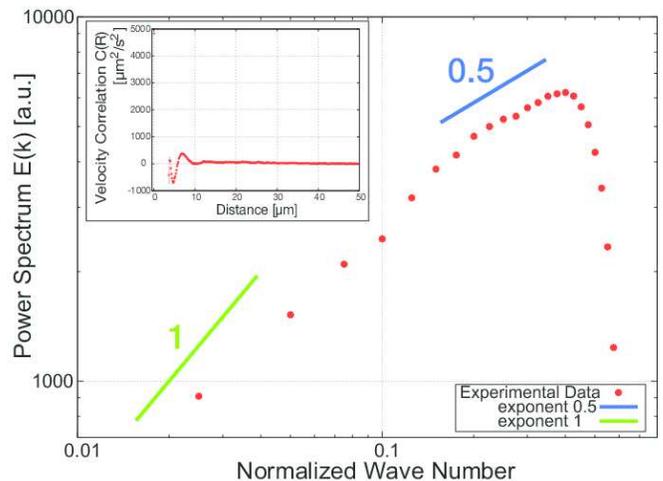}
 \end{center}
 \caption{\label{VeloCorr} (Color online) Log-log plot of the power spectral density of the velocity field of the Janus particles.
The power spectrum was calculated according to \equref{SpectrumEq} using the experimentally obtained velocity correlation function defined as \equref{CorrR}. The velocity correlation function $C(R)$ is shown in the inset. The blue and green lines are slopes with the exponents 0.5 and 1 respectively to guide the eye. Wave number is normalized by the wave number corresponding to the diameter of the Janus particle. The vertical axis is not normalized.}
\end{figure}

The power spectral density $E(k)$ of the two-dimensional velocity field of the Janus particles was obtained by applying two-dimensional Fourier transformation to a velocity correlation function $C(\bm{R})$,
\begin{eqnarray}
&&C(\bm{R}) := \langle \bm{v}(t, \bm{r}) \cdot \bm{v}(t, \bm{r}+\bm{R}) \rangle_{t, \bm{r}} \ \label{CorrR} ,\\
&&E(k) = 2\pi k \int \frac{\mathrm{d}^2 \bm{R}}{(2 \pi)^2} \  \mathrm{e}^{-i\bm{k}\cdot\bm{R}} \langle \bm{v}(t, \bm{r}) \cdot \bm{v}(t, \bm{r}+\bm{R}) \rangle_{t, \bm{r}}\ ,\;\;\; \label{SpectrumEq}
\end{eqnarray}
where $\bm{v}(t,\bm{r})$ is the velocity of the particle at the position $\bm{r}$ at time $t$, $\bm{R}$ is the relative position vector from one particle to the other, $\bm{k}$ is the wave vector in the Fourier space, and $k$ is the magnitude of $\bm{k}$. All the vectors are two-dimensional. The angle brackets mean the average over all the particle pairs at relative position $\bm{R}$ at every time step. The number of calculated pairs was in total $1.2\times 10^8$. Because our system is isotropic, the velocity correlation function $C(\bm{R})$ does not depend on the argument of $\bm{R}$ and we can replace $C(\bm{R})$ with $C(R)$, where $R$ is the magnitude of $\bm{R}$ (the inset of \figref{VeloCorr}).
Because the distance between the centroids of the particles must be larger than their diameters due to exclusive interactions, the experimentally obtained velocity correlation function has no data points at the distance between 0 \micron \ and 3 \micron. We interpolated the velocity correlation function and obtained the power spectrum according to \equref{SpectrumEq} (\figref{VeloCorr}) \cite{InterpolationMethod}. The wave numbers in \figref{SpectrumEq} is normalized by the wave number $2\pi /(3\, \mathrm{\mu m})$, which corresponds to the particle diameter. The actual calculation procedure is summarized in \Appref{AppendixSpectrumMethod}.

The obtained power spectrum has a broad peak with a lower slope from 0.2 to 0.4 in terms of the normalized wave number, which corresponds to the spatial scales from 7 \micron \ to 15 \micron. As the wave number goes smaller, the slope of the power spectrum gradually changes roughly from the exponent 0.5 to 1. The exponent 1 in the smaller wave number regime is trivial because the Fourier transform of the correlation function cannot detect large wave length fluctuations sufficiently and the integrals in \equref{SpectrumEq} and \equref{SpecCalcEq} behave as constants with respect to $k$, which results in $E(k) \propto k$. In bacterial turbulence \cite{Wensink2012}, the power spectrum with the exponent $5/3$ was suggested, but it is inappropriate to compare the forms of their power spectra precisely due to the limited observed ranges of the wave numbers for both cases.
Furthermore, in the study on agent-based model with short-range polar alignment and anti-parallel alignment at larger distances by Gro\ss mann et al. \cite{Grossmann2014,Grossmann2015}, it was demonstrated that the scalings in the power spectra depend on the choice of interaction parameters and thus the exponents and the form of the power spectra are not universal.

\subsection{Velocity correlation in detail}
We look into the velocity correlation in detail to investigate the origins of the form of the power spectrum. In the inset of \figref{VeloCorr}, the oscillation of the velocity correlation function $C(R)$ is observed, indicating the existence of structures with finite length scales. This form of $C(R)$ is intriguing because Gro\ss mann et al. also observed quite similar oscillation in $C(R)$ of the turbulent states of their model \cite{Grossmann2014}. To quantify the angular information, which was dismissed in $C(R)$, we define another velocity correlation function $C'(R, \psi)$,
\begin{eqnarray}
C'(R, \psi) :=\left\langle \bm{v}(t, \bm{r}) \cdot \bm{v}(t, \bm{r}+R \bm{\hat{e}}_\psi)\right\rangle_{t, \bm{r}},\  \ \ \ && \label{CorrPsi}
\end{eqnarray}
where $\bm{\hat{e}}_\psi$ is a unit vector in the direction of $\psi+\arg{\bm{v}(t, \bm{r})}$. $C'(R, \psi)$ can extract the information on how the other particles are moving from a viewpoint of a reference particle.

\begin{figure}[htbp]
 \begin{center}
  \subfigure[Colormap of the velocity correlation function $C'(R, \psi)$]{\includegraphics[width=0.8\columnwidth]{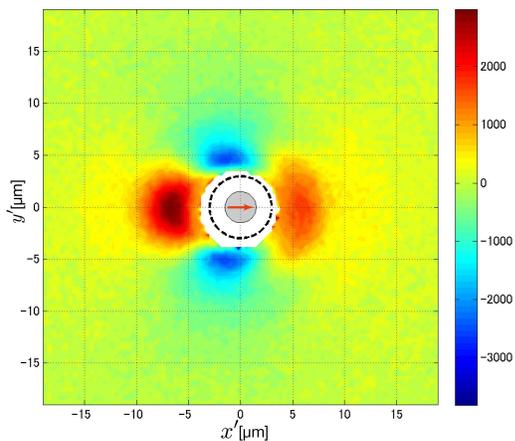}}
  \subfigure[Velocity correlation in $x'$ or $y'$ direction]{\includegraphics[width=0.8\columnwidth]{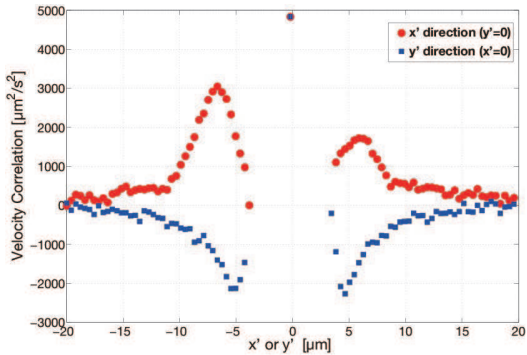}}
 \end{center}
  \caption{\label{FigVelocityCorrAngle} (Color online) (a) Colormap of the velocity correlation function $C'(R, \psi)$ defined in \equref{CorrPsi}. The coordinate is $x'=R\cos{\psi}$ and $y'=R\sin{\psi}$. Note that $x'$ and $y'$ are different from the axes defined in \figref{ExpSetup}. The filled gray circle represents a reference Janus particle, whose direction of motion (the red arrow inside) is rearranged to the right ($x'$ direction). The dashed circle corresponds to twice the radius, which the other particle cannot enter due to exclusive interaction. In reality, repulsive electrostatic interactions, surfactants absorbed on the colloidal surface, and the suspending fluid dominated by Stokesian dynamics keep the particles apart and they cannot get close to this dashed circle. No particles are detected in the white region. (b) Slices of (a) at $y'=0$ ($x'$ direction, parallel to the velocity, red circles) and $x'=0$ ($y'$ direction, perpendicular to the velocity, blue squares). Negative correlation in $y'$ direction shows anti-parallel alignment and positive correlation in $x'$ direction indicates that the particles are following other particles in front of them.}
\end{figure}

The velocity correlation function $C'(R, \psi)$ shown in \figref{FigVelocityCorrAngle} indicates that anti-parallel alignment in the transverse directions and polar alignment in the longitudinal directions coexist. The negative correlation in $y'$ ($\psi=\pm\pi/2$) direction, which is perpendicular to the direction of motion, indicates that the particles next to each other tend to move in the opposite directions. This anti-parallel alignment can be partially explained by considering hydrodynamic interactions between the Janus particles in a sense that the parallel configuration of two Janus particles moving in the same direction is unstable (see \Appref{AppendixHydrodynamicInteraction} for a detailed explanation). The positive correlation in $x'$ direction, which is parallel to the direction of motion, indicates that the particles are following other particles in front of them. In the study by Gro\ss mann et al. \cite{Grossmann2014,Grossmann2015}, it was shown that the coexistence of polar and anti-parallel alignment lead to effective negative viscosity and can trigger active turbulence. This coexistence might cause effective negative viscosity in suspensions of the Janus particles and account for the turbulent state in the Janus particles system, although the angular dependence of the alignment interactions is different from that of their agent-based model.

\subsection{Local order parameters}
\begin{figure}[hbtp]
 \begin{center}
  \includegraphics[width=1\columnwidth]{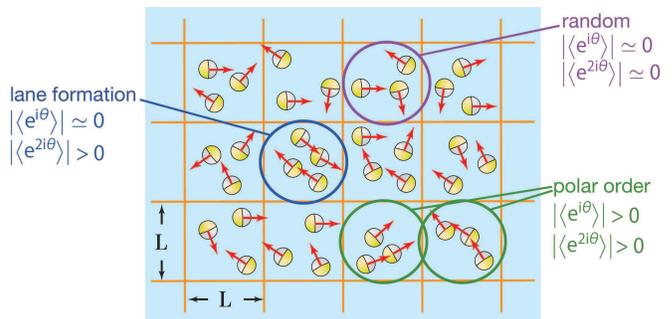}
 \end{center}
  \caption{\label{FigOrderParametersMeaning} (Color online) The local average values of the orientational order parameter $\abs{\langle \mathrm{e}^{\mathrm{i}\theta} \rangle}$ and the nematic order parameter $\abs{\langle \mathrm{e}^{2\mathrm{i}\theta} \rangle}$ can distinguish local states of collective behaviors.}
\end{figure}

\begin{figure}[hbtp]
 \begin{center}
  \subfigure[Probability distribution of $\abs{\langle \mathrm{e}^{\mathrm{i}\theta} \rangle}$]{\includegraphics[width=0.9\columnwidth]{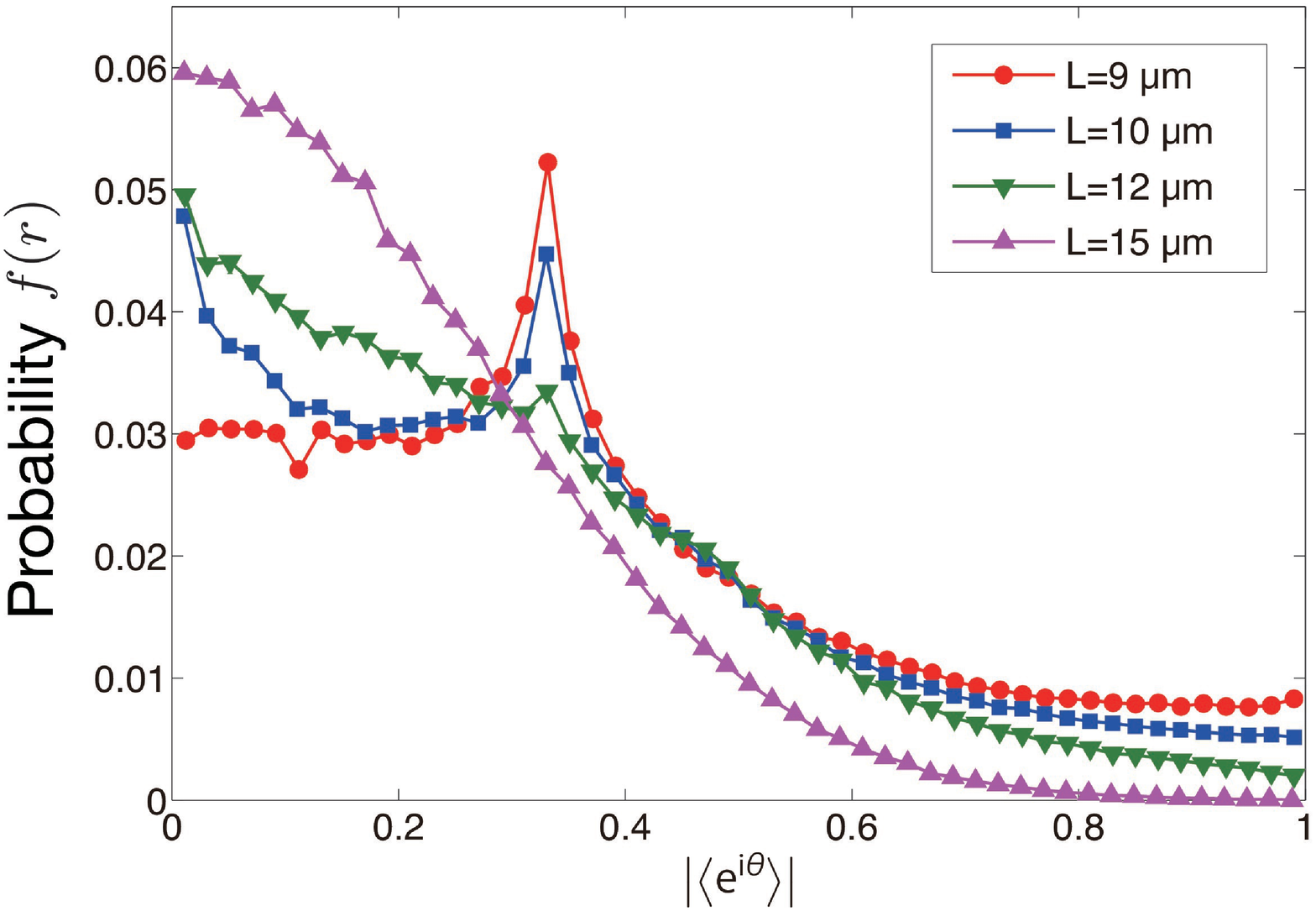}}
  \subfigure[Probability distribution of $\abs{\langle \mathrm{e}^{2\mathrm{i}\theta} \rangle}$]{\includegraphics[width=0.9\columnwidth]{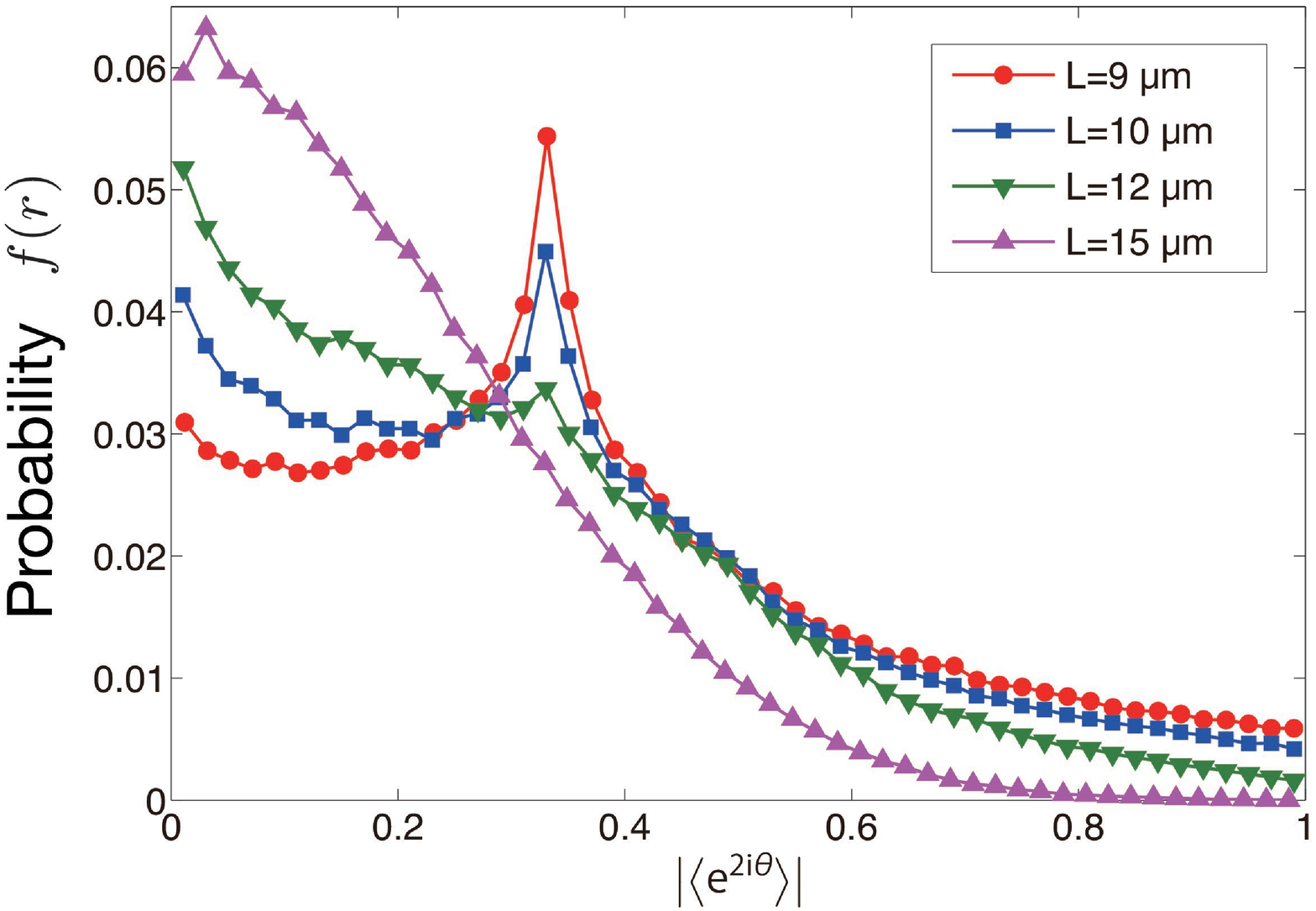}}
 \end{center}
  \caption{\label{FigOrderParametersHist} (Color online) The probability distributions of the order parameters defined as \equref{EqProbDist} are calculated at the scale of $L=9, 10, 12, 15 \ \mathrm{\mu m}$ respectively. When $L$ is smaller than 12 \micron, both order parameters have non-zero peaks meaning that the particles are locally moving in the same direction but opposing lanes are not formed. These peaks gradually grow as $L$ is decreased. In the calculation of order parameters, the regions in the grid including fewer than three particles are neglected in order to extract the results coming from the collective dynamics.}
\end{figure}

In order to quantify structures and orders emerging from the two-body velocity correlations which we have seen in the previous subsection, we calculate two locally defined order parameters: an orientational order parameter $\abs{\langle \mathrm{e}^{\mathrm{i}\theta} \rangle}$ and a nematic order parameter $\abs{\langle \mathrm{e}^{2\mathrm{i}\theta} \rangle}$, where $\theta$ represents the direction of instantaneous velocity of each Janus particle (\figref{FigOrderParametersMeaning}).  These order parameters can clarify whether nematic ordering or lane formation is observed as they are reported in numerical studies with anti-parallel alignment interactions \cite{Wensink2012, Wensink2012b, Ginelli2010}. We calculate these order parameters locally in a $L\times L$ grid to make histograms of their values. The grid size $L$ is varied from 9 \micron \ to 15 \micron. The obtained probability distributions are shown in \figref{FigOrderParametersHist}. $f(r)$ plotted in \figref{FigOrderParametersHist} are probability distribution functions (pdf) with respect to moduli of $\langle \mathrm{e}^{\mathrm{i}\theta} \rangle$ and $\langle \mathrm{e}^{2\mathrm{i}\theta} \rangle$,
\begin{flalign}
\begin{split}
f\left(\abs{\langle \mathrm{e}^{\kappa\mathrm{i}\theta} \rangle}=r\right)\propto \frac{1}{2\pi r} \int_0^{2\pi} \mathrm{d}\phi \ \mathrm{pdf}\left(\abs{\langle \mathrm{e}^{\kappa\mathrm{i}\theta} \rangle}=r, \right.& \\
\left. \arg{\langle \mathrm{e}^{\kappa\mathrm{i}\theta} \rangle}=\phi\right)&,
\end{split}
\label{EqProbDist}
\end{flalign}
where $\kappa=1$ or $2$ for the orientational order parameter or the nematic order parameter respectively, and $\arg$ is a function returning an argument of a complex number.

\begin{figure}[hbtp]
 \begin{center}
  \includegraphics[width=\columnwidth]{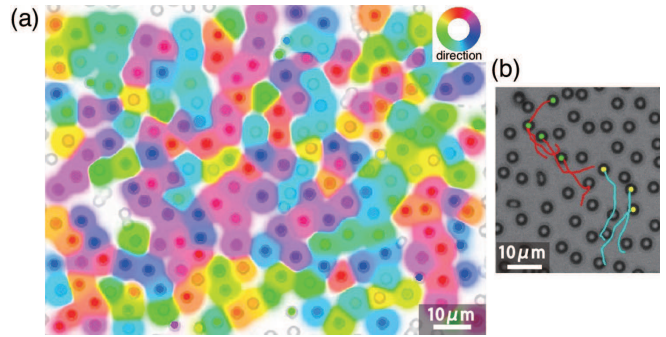}
 \end{center}
  \caption{\label{FigCluster} (Color online) (a) In a snapshot of the experimental movie, each particle is colored depending on its direction of instantaneous motion. Spaces between the particles are also colored according to average velocities of the nearby particles with Gaussian weight (see \Appref{AppendixVorticity} for detail). We can see some regions connected with same colors \cite{MovieCluster}. (b)Trajectories of the particles moving in the same direction are overlaid on the snapshot at the last moment. Different colors indicate different clusters \cite{MovieLaning}. The scale bars correspond to 10\ \micron .}
\end{figure}

As the grid size is decreased, non-zero peaks gradually appear and grow higher for both order parameters. This indicates that the particles are locally moving in the same direction forming structures like trains but the opposing lanes are not formed. The particles exhibit local orientational order, although they do not have global orientational order, nematic order, or positional order (see \Appref{AppendixPairCorrelation}). When we examine the captured movies carefully, some local clusters moving in the same direction for a while can be found frequently ( \figref{FigCluster} and supplemental movies \cite{MovieLaning, MovieCluster}). The fact that the non-zero peaks for the two order parameters are observed in various grid sizes suggests that there exist various scales of these clusters up to 12 \micron \ in size. These structures are formed and broken up at all times, changing their cluster sizes. 

The length scales where the broad peak is observed in \figref{VeloCorr} corresponds to those where the non-zero peaks are observed for  the two order parameters in \figref{FigOrderParametersHist}. Furthermore, the strongest positive correlation in the longitudinal direction in \figref{FigVelocityCorrAngle} at $x=-7 \,\mathrm{\mu m}$ corresponds to the highest position of the broad peak in the power spectrum. Thus we conclude that the formation of clusters of particles moving in the same direction is a possible cause of the origin of the broad peak found in the power spectrum of the velocity field of the Janus particles.

\section{\label{discussion}Discussion}
In the usual fluid turbulence, according to Kolmogorov's law \cite{LandauFluidMech}, the exponent of the power law observed in the power spectrum is negative, while the slope in our experiment on the Janus particles is positive. The crucial difference is in the spatial scales where energy is injected into the systems. In the usual fluid turbulence, energy is continuously injected externally by boundary conditions such as pressure gradient and shear stress. Consequently, the energy input occurs at the scale of its system size, which corresponds to the smallest wave number in the spectrum, and this energy is transferred to the larger wave number regime, finally dissipating into heat at small eddies. In contrast, in the active turbulence of the Janus particles or bacteria \cite{Wensink2012}, energy is injected in the bulk by the particles themselves, whose length scale is the smallest in the system. This means that energy input is occurring at the largest wave number in the power spectrum. According to our experimental results showing that the Janus particles are forming clusters with the particles moving in the same directions, it is possible that some conserved quantities, which are energy and enstrophy in the case of fluid turbulence, are cascading from the particle scale to the larger structures. 

Since the beginning of studies on active matter, most effort has been put on describing macroscopically coherent states or transitions between ordered states and disordered states. However, to the best of our knowledge, there are no experimental systems of microswimmers except for the Quincke rollers that exhibit macroscopically directed ordered phase, which is often obtained in many numerical studies including Vicsek model  \cite{Vicsek1995, Bricard2013}. It is of great importance to understand from the microscopic point of view why it is difficult for microswimmers to align in the same direction.

The interactions dominating the collective dynamics of self-propelled colloidal particles under an electric field are both hydrodynamic and electrostatic ones. In the case of the Quincke rollers, their electrostatic interaction had been already understood, so the calculations considering the hydrodynamic interaction clearly explained the mechanism of the transition to the coherent motion \cite{Bricard2013}. However, as is mostly the case for theoretical calculations, their calculation on the hydrodynamic interactions relies on far-field approximation, in which each microswimmer is represented as a point singularity \cite{Brotto2013}.

Near-field hydrodynamic interactions are much more important to understand collective behaviors of swimmers than far-field ones, especially in the case of spherical particles moving very close to a surface such as our Janus particles. Because spherical particles do not have strong steric alignment interactions, hydrodynamic interactions dominate their dynamics. When the particles are moving very close to a surface or confined between two walls, far-field flow is strongly suppressed and only near-field interactions play an important role in their dynamics \cite{Blake1974,Liron1976}. When we neglect the near-field interactions and the electrostatic interactions, the remaining far-field interactions cannot account for the difference in the macroscopic behaviors of the Janus particles and the Quincke rollers. Their far-field interactions are the same because the two electrodes close to the particles damp all the singularities of flow except source dipoles created by each particle \cite{Liron1976,Brotto2013}. Therefore, the difference in their macroscopic states should be explained in terms of their near-field hydrodynamic interactions or electrostatic interactions, which must be very different between the two due to the difference in the ways how their symmetries are broken and the direction of self-propulsion is determined.

Because of the difficulties in the theoretical calculations on near-field hydrodynamics, numerical simulations have been performed to understand the hydrodynamics of swimmers \cite{Ishikawa2008,Molina2013}. In these studies, microswimmers are represented as squirmers, which have spherical bodies with slip boundary conditions \cite{Blake1971}. This simplified model is useful for studying our Janus particles because their slip velocity is theoretically calculated in the low frequency limit \cite{Squires2004} (see \Appref{AppendixHydrodynamicInteraction}).  Numerical simulations on Janus squirmers with electrostatic interactions are now in progress \cite{Delfau}.

\section{Summary}
To study turbulent states of active particles by using non-living microswimmers, we realized an experimental system of Janus particles propelling in a quasi-two-dimensional plane with high density and high motilities. The area fraction of the particles can be set as high as 0.24, which has not been achieved because of aggregation of the particles in the previous studies \cite{Suzuki2011}. Their average self-propelling speed reaches 61.4 \micron/s, which is about three times faster than usual bacteria or the Janus particles in the previous studies \cite{Gangwal2008,Suzuki2011,Boymelgreen2014}. We tracked all the particles in the observed area and calculated the power spectrum of their velocity field which shows the broad peak ranging over a length scale of several times the particle size. 

Calculation of the velocity correlation functions revealed the coexistence of the polar alignment and the anti-parallel alignment, which had been demonstrated to be responsible for the effective negative viscosity of active suspensions and the emergence of active turbulence \cite{Grossmann2014,Grossmann2015}. However, it remains to be elucidated how these alignment effects arise from the microscopic interactions such as the hydrodynamic and the electrostatic interactions.

The probability distributions of the order parameters indicated the formation of local clusters moving in the same direction. Judging from the correspondence between the spatial scales where the broad peak is observed and those where the clusters are found, the cluster formation can be understood as one of the possible causes of the power law.

In future, we shall focus on measurements of the properties of their dynamics by changing control parameters such as the amplitude and the frequency of the applied AC electric field, and the number density of the particles.
In this system the particles tend to aggregate at smaller amplitude of applied voltage, so we have not successfully analyzed how the collective dynamics of our Janus particles depend on the amplitude and the frequency of the electric field. Because we can tune the interactions, the self-propulsion speed, and the direction of motion of the Janus particles by controlling the amplitude and the frequency of the electric field \cite{Suzuki2011}, it shall serve as an ideal experimental setup to figure out some universal laws of collective motion of self-propelled particles from their microscopic interactions.

\section*{Acknowledgments}
We are grateful to Hong-Ren Jiang, Jean-Baptiste Delfau, Ken H. Nagai and Ryo Suzuki for fruitful discussions. We thank H. P. Zhang for providing his MATLAB codes for detecting bacteria. We also thank Kazumasa A. Takeuchi and Kyogo Kawaguchi for their critical reading of the manuscript.
This work is supported by Grant-in-Aid for Japan Society for Promotion of Science (JSPS) Fellows (Grant Number 26$\cdot$9915), the JSPS Core-to-Core Program ``Non-equilibrium dynamics of soft matter and information,'' and KAKENHI (No. 25103004, ``Fluctuation \& Structure'') from MEXT, Japan.

\appendix
\section{\label{AppendixVorticity}Vorticity map}
\begin{figure}[hbtp]
 \begin{center}
  \includegraphics[width=\columnwidth]{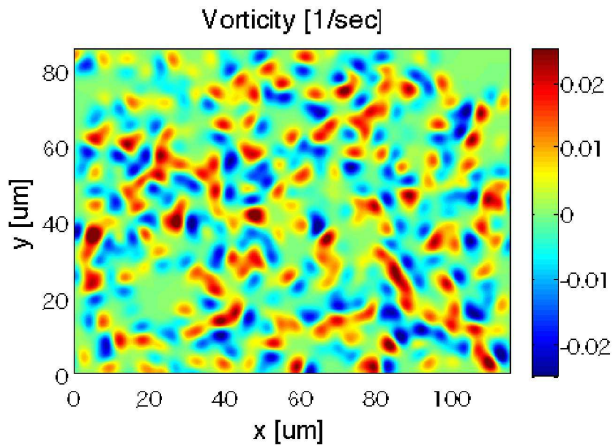}
 \end{center}
 \caption{\label{FigVorticityColorPlot}(Color online) Color map of the vorticity. Various scales of vortices with positive and negative vorticity coexist.}
\end{figure}

In order to examine how turbulent the velocity field of the Janus particles is, we make a color map of the vorticity of the velocity field (\figref{FigVorticityColorPlot}). The velocity of each particle is calculated from its trajectory, and then the time series of velocities are smoothed by taking moving average over 21 frames. Because the velocity is originally defined only on the centroids of the particles, we apply Gaussian filter to obtain the velocity field. The standard deviation of the applied Gaussian filter is chosen to be the radius of the Janus particles. Various scales of vortices with positive and negative vorticity coexist and evolve in time \cite{MovieVorticity}.

\section{\label{AppendixSpectrumMethod}Calculation of Spectrum}
When we consider an isotropic two-dimensional system, the velocity correlation function defined as \equref{CorrR} is independent of the angle of $\bm{R}$, leading to
\begin{eqnarray}
C(R) &=& \langle \bm{v}(t, \bm{r}) \cdot \bm{v}(t, \bm{r}+\bm{R}) \rangle_{t, \bm{r}}.
\end{eqnarray}
Calculation of its power spectral density defined in \equref{SpectrumEq} is the following:
\begin{eqnarray}
E(k) &=& \frac{k}{2\pi} \int \mathrm{d}^2 \bm{R} \  \mathrm{e}^{-i\bm{k}\cdot\bm{R}} C(R) \\
&=&  \frac{k}{2\pi} \int_0^{\infty} R \mathrm{d} R \int_0^{2\pi} \mathrm{d}\phi \  \mathrm{e}^{-i\bm{k}\cdot\bm{R}} C(R) ,\label{Rewrite}
\end{eqnarray}
where $\phi$ is the argument of $\bm{R}$. We can calculate the part of the integrand with $\phi$ by using the periodicity of cosine function,
\begin{eqnarray}
\int_0^{2\pi} \mathrm{d}\phi \  \mathrm{e}^{-i\bm{k}\cdot\bm{R}} &=& \int_0^{2\pi} \mathrm{d}\phi \  \mathrm{e}^{-i kR\cos{(\phi-\alpha)}} \\
&=&  \int_0^{2\pi} \mathrm{d}\phi \  \mathrm{e}^{-i kR\cos{\phi}} \\
&=&  2\pi J_0(kR) ,\label{BesselThetaDependent}
\end{eqnarray}
where $\alpha$ is the argument of $\bm{k}$ and $J_0$ is the 0th order Bessel function of the first kind. By substituting \equref{BesselThetaDependent} into \equref{Rewrite}, we obtain the final result,
\begin{eqnarray}
E(k) &=& k \int_0^{\infty} \mathrm{d} R \  C(R) \cdot R \  J_0(kR) . \label{SpecCalcEq}
\end{eqnarray}
We can calculate the power spectrum of an isotropic velocity field from its correlation function by using  \equref{SpecCalcEq}.

\section{\label{AppendixHydrodynamicInteraction}Flow field and hydrodynamic interaction}
Janus particles are interacting with each other through hydrodynamic and electrostatic interactions. The electrostatic interaction is quite complicated, because electrical double layers around the particles have their own characteristic charging time and thus their thickness depends on the frequency of the applied AC electric field. On the other hand, the hydrodynamic interaction can be understood by solving the Stokes equation. From the hydrodynamic interactions, we can conclude that the parallel configuration of two Janus particles moving in the same direction is unstable.
 
We can solve the Stokes equation to obtain the flow field around a Janus particle by assuming slip velocity on the particle, if we neglect the nearby electrodes and only consider one Janus particle suspending in fluid.
The flow around the Janus particle is dominated by Stokesian dynamics, because the Reynolds number of the flow is as low as $10^{-4}$. On the surface of the Janus particle, the fluid is driven by ICEO. The ICEO flow on the polystyrene side is considerably weaker than that on the metal side, so it is justified to neglect the flow on the polystyrene side \cite{Squires2006}. We regard that this ICEO flow exists only in the vicinity of the surface, which is the thin electric double layer with the thickness of its Debye length \cite{Squires2004}. We replace the non-slip boundary condition on the metal hemisphere to the slip boundary condition with the slip velocity calculated in a theoretical study \cite{Squires2006}.

\begin{figure}[hbtp]
 \begin{center}
  \includegraphics[width=0.6\columnwidth]{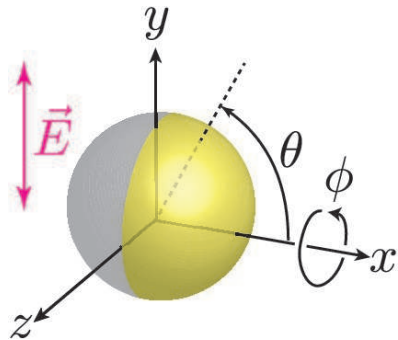}
 \end{center}
 \caption{\label{FigSphericalCoordinate}(Color online) The spherical coordinate used for calculation. The origin is at the center of the Janus particle. The right hemisphere ($x>0$, or $|\theta|>\pi/2$) is covered with metal. Note that the axes defined in this \Appref{AppendixHydrodynamicInteraction} are different from those used in Figs. \ref{ExpSetup}, \ref{FigVelocityCorrAngle}, \ref{FigVorticityColorPlot}.}
\end{figure}

\begin{figure}[hbtp]
 \begin{center}
  \includegraphics[width=0.9\columnwidth]{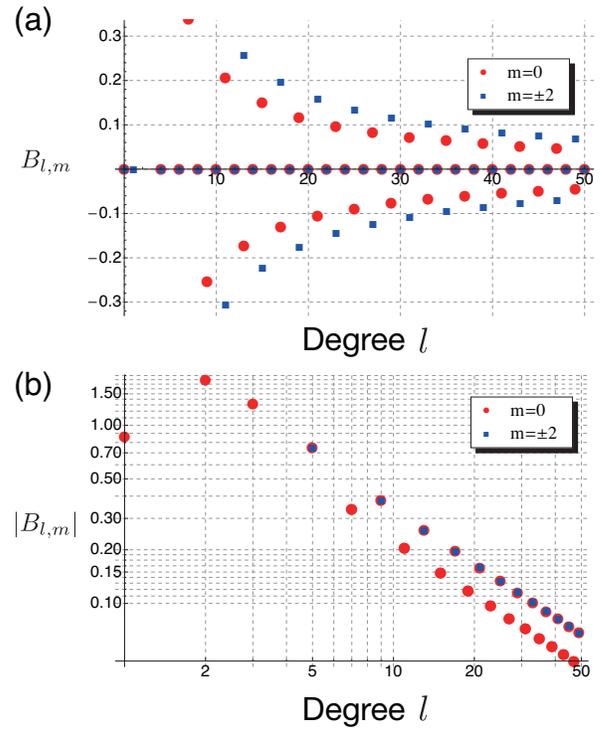}
 \end{center}
 \caption{\label{FigSphericalHarmonics}(Color online) Values of the coefficients of spherical harmonics (a) $B_{l,m}$ in a linear plot and (b) $\left| B_{l,m} \right|$ in a log-log plot. $B_{l,m}$ has non-zero values only when $m=0, \pm2$. For any $l$, $B_{l,+2}=B_{l,-2}$ holds due to symmetry in the sign of $\phi$.}
\end{figure}

The method to solve the equation is the following \cite{HappelBrenner}. First of all, we consider the particle frame, in which the particle is at rest. We take a spherical coordinate as depicted in \figref{FigSphericalCoordinate}.
The slip velocity on the surface of the Janus particle $\bm{u}_s$ is given as
\begin{align}
\bm{u}_s(\theta, \phi)  = \hspace{6.5cm} \nonumber \\
\left\{ \begin{array}{ll}
    -\frac{9}{8}U_0 (\sin{2\theta}\cos^2{\phi}\:\hat{\bm{\theta}}-\sin{\theta}\sin{2\phi}\:\hat{\bm{\phi}}) & (|\theta|<\pi/2) \\
    0 & (|\theta|\geq\pi/2), \label{EqSlipVelocity}
  \end{array} \right.
\end{align}
where $U_0$ is the typical speed of ICEO flow, and $\hat{\bm{\theta}}$ and $\hat{\bm{\phi}}$ are the unit vectors with $+\theta$ and $+\phi$ directions respectively.
In order to solve the Stokes equation, we need another boundary condition, which is the velocity at infinity $\bm{u}_\infty$. Because we are in the co-moving frame with the particle, the velocity at infinity has the opposite sign and the same amplitude of the velocity of the particle $\bm{U}_\mathrm{ICEP}$ in the laboratory frame. $\bm{U}_\mathrm{ICEP}$ can be calculated from the slip velocity \cite{Squires2006} and it leads to
\begin{align}
\bm{u}_\infty &= - \bm{U}_\mathrm{ICEP} = \frac{1}{4\pi}\int\bm{u}_s \mathrm{d}\Omega =  \frac{9}{64} U_0 \hat{\bm{x}},
\label{EqInfinityVelocity}
\end{align}
where $\bm{U}_\mathrm{ICEP}$ is the velocity of the particle in the laboratory frame, $\mathrm{d}\Omega$ is an element of solid angle, and $\hat{\bm{x}}$ is a unit vector in the direction of $x$ axis.
We expand the boundary conditions on the surface with spherical harmonics $Y_l^m(\theta, \phi)$, 
\begin{align}
\left. \frac{\bm{r}}{r}\cdot (\bm{u}-\bm{u}_\infty) \right|_{r=a} &=\sum_{l=0}^\infty \sum_{m=-l}^l A_{l,m} Y_l^m(\theta ,\phi )\\
\left. -r \nabla \cdot (\bm{u}-\bm{u}_\infty) \right|_{r=a} &=\sum_{l=0}^\infty \sum_{m=-l}^l B_{l,m} Y_l^m(\theta ,\phi ) \\
\left. \bm{r}\cdot \nabla \times (\bm{u}-\bm{u}_\infty) \right|_{r=a} &=\sum_{l=0}^\infty \sum_{m=-l}^l C_{l,m} Y_l^m(\theta ,\phi )  
\end{align}
where $\bm{r}$ is a position vector in the spherical coordinate, $r$ is the magnitude of $\bm{r}$, $a$ is the radius of the Janus particle, $Y_l^m(\theta, \phi)$ is a spherical harmonic function of degree $l$ and order $m$, and $A_{l,m}$, $B_{l,m}$, and $C_{l,m}$ are coefficients of expansion. By using \equref{EqSlipVelocity}, \equref{EqInfinityVelocity} and an orthogonality of spherical harmonics, we can obtain the coefficients as
\begin{align}
A_{l,m} &= \int_0^{2\pi} \mathrm{d} \phi \int_0^{\pi} \mathrm{d} \theta \left[ \left. \frac{\bm{r}}{r}\cdot (\bm{u}-\bm{u}_\infty) \right|_{r=a} \right] Y_l^m(\theta, \phi)\\
&= \int_0^{2\pi} \mathrm{d} \phi \int_0^{\frac{\pi}{2}} \mathrm{d} \theta \left[ -\frac{9}{64} U_0 \cos{\theta} \right] Y_l^m(\theta, \phi)
\end{align}
\begin{align}
B_{l,m} = \int_0^{2\pi} \mathrm{d} \phi \int_0^{\pi} \mathrm{d} \theta \left[\left. -r \nabla \cdot (\bm{u}-\bm{u}_\infty) \right|_{r=a} \right] Y_l^m(\theta, \phi)&\\
= \int_0^{2\pi} \mathrm{d} \phi \int_0^{\frac{\pi}{2}} \mathrm{d} \theta \left\{ \frac{9}{4} U_0 \left[ (2\cos^2{\theta}-\sin^2{\theta})\cos^2{\phi} \right. \right. & \nonumber \\
\left. \left. -\cos{2\phi} \right]  \right\} Y_l^m(\theta, \phi) &
\end{align}
\begin{align}
C_{l,m} = & \int_0^{2\pi} \mathrm{d} \phi \int_0^{\pi} \mathrm{d} \theta \left[ \left. \bm{r}\cdot \nabla \times (\bm{u}-\bm{u}_\infty) \right|_{r=a} \right] Y_l^m(\theta, \phi) \\
= & \int_0^{2\pi} \mathrm{d} \phi \int_0^{\frac{\pi}{2}} \mathrm{d} \theta  \cdot 0 \cdot  Y_l^m(\theta, \phi) \\
= & 0.
\end{align}
$A_{l,m}$ has a non-zero value only at $A_{1,0}=-\frac{9}{64}\cdot 2\sqrt{\frac{\pi}{3}}$. Absolute values of non-zero $B_{l,m}$ are plotted in \figref{FigSphericalHarmonics}, which decay algebraically with the increase of $l$.

Because the expansions are infinite series, we take up to $l=50$ terms to calculate the flow field. Judging from the values of $B_{2,0}=1.78$ and $B_{49,0}=-0.0456$, truncation error can be estimated at approximately less than 3 \%.

By defining six kinds of harmonics,
\begin{align}
X(l,\theta ,\phi) &:= \sum_{m=-l}^l A_{l,m} Y_l^m(\theta ,\phi )\\
Y(l,\theta ,\phi) &:= \sum_{m=-l}^l B_{l,m} Y_l^m(\theta ,\phi ) \\
Z(l,\theta ,\phi) &:= \sum_{m=-l}^l C_{l,m} Y_l^m(\theta ,\phi )\\
p_{-(n+1)} &:= \frac{\eta  (2 n-1)}{n+1} \frac{1}{a} \left(\frac{a}{r}\right)^{n+1} \nonumber \\ 
 &\:\:\:\:\:\: \times \left[ (n+2)X(n,\theta ,\phi)+Y(n,\theta ,\phi ) \right]\\
\Phi_{-(n+1)} &:= \frac{1}{2(n+1)} a \left(\frac{a}{r}\right)^{n+1} \left[ nX(n,\theta ,\phi)+Y(n,\theta ,\phi ) \right]\\
\chi_{-(n+1)} &:= \frac{1}{n(n+1)}  \left(\frac{a}{r}\right)^{n+1} Z(n,\theta ,\phi),
\end{align}
we can obtain the flow field as
\begin{align}
\bm{u}-&\bm{u}_\infty = \sum_{n=1}^\infty \left[ \nabla \times (\bm{r} \chi_{-(n+1)}) + \nabla \Phi_{-(n+1)} \right. \nonumber \\
&\:\:\: \left. - \frac{n-2}{\eta 2 n (2n-1)} r^2 \nabla p_{-(n+1)} + \frac{n+1}{\eta n (2n-1)} \bm{r} p_{-(n+1)} \right],
\end{align}
where $\eta$ is a shear viscosity of the fluid. The viscosity $\eta$ cancels out and does not appear in the final expression of $\bm{u}$, while it does appear in the pressure formula \cite{PressureRepresentation}.

\begin{figure}[hbtp]
 \begin{center}
  \includegraphics[width=\columnwidth]{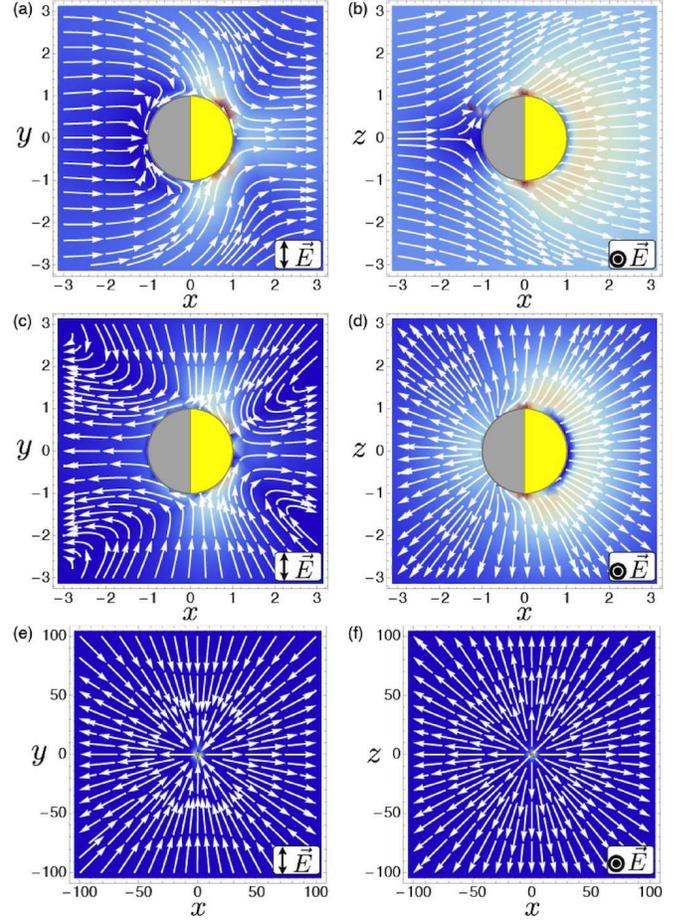}
 \end{center}
 \caption{\label{FigFlowSet}(Color online) Streamlines of the flow field around a Janus particle. (a)(c)(e) Flow parallel to the electric field. (b)(d)(f) Flow perpendicular to the electric field. (a) and (b) are the flow field in the particle frame, and (c)-(f) are that in the laboratory frame. The far-field flow shown in (e) is reminiscent of that of pusher-type microswimmers. The radius of the particle is set to 1. The color indicates the amplitude of the flow field. The speed of the flow is faster in the red region.}
\end{figure}

The solved flow field is shown in \figref{FigFlowSet}. We plot the flow field in the particle frame $\bm{u}$ and that in the laboratory frame $\bm{u}-\bm{u}_\infty$. The far-field flow in the plane parallel to the electric field corresponds to the flow field created by pushers \cite{Molina2013}, because the higher order terms decay much faster than the lowest order term representing a force dipole. Because the flow in this plane converges in $y$ direction, the flow in the plane perpendicular to the electric field is divergent.

\begin{figure}[hbtp]
 \begin{center}
  \includegraphics[width=0.8\columnwidth]{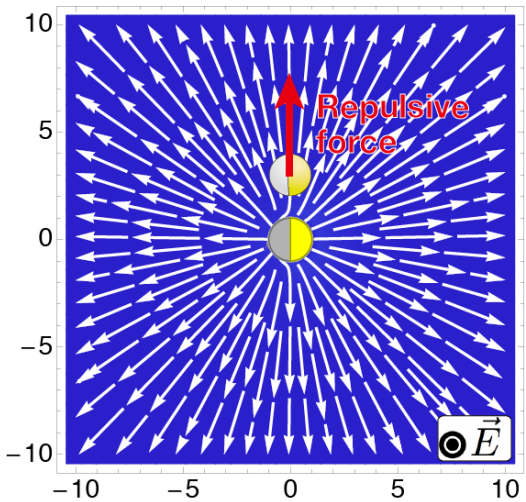}
 \end{center}
 \caption{\label{FigParallelInstability}(Color online)  The parallel configuration of the two Janus particles moving in the same direction is unstable. Because the observed plane in the experiment is the plane perpendicular to the applied electric field, the flow shown in this figure is accountable for the hydrodynamic interaction. The hydrodynamic flow between the two particles repels each other apart. The radius of the particle is set to 1.}
\end{figure}

We can finally understand the reason of the instability of why the parallel configuration of the two Janus particles moving in the same direction is unstable (\figref{FigParallelInstability}). The hydrodynamic flow created by one particle repels the other particle. This interaction makes this configuration unstable and the particles push each other to go apart. These kinds of instabilities have been discussed especially for the case of pusher-type microswimmers in some references \cite{Subramanian2009,Tsang2014}.

Here we do not take into account the effects of the walls or the electrodes, and we cannot explain why the particles tend to form clusters moving in the same direction. Precise calculations or hydrodynamic simulations with the walls and the electrostatic interactions might account for the cluster formation. 

\section{\label{AppendixPairCorrelation}Pair correlation function}
In order to examine the positional order of the crowd of the Janus particles, we calculate a pair correlation function $g(r)$ defined as
\begin{eqnarray}
g(r) &:=& \frac{1}{2\pi r \rho_0} \left\langle \frac{1}{N} \sum_{i\neq j} \delta{(r-r_{ij})} \right\rangle,
\label{EqPairCorr}
\end{eqnarray}
where $\rho_0$ is the average number density, $N$ is the total number of the particles, and $r_{ij}$ is the distance between the $i$-th and the $j$-th particles. The brackets mean average over all the time series.

\begin{figure}[htbp]
 \begin{center}
  \includegraphics[width=83mm]{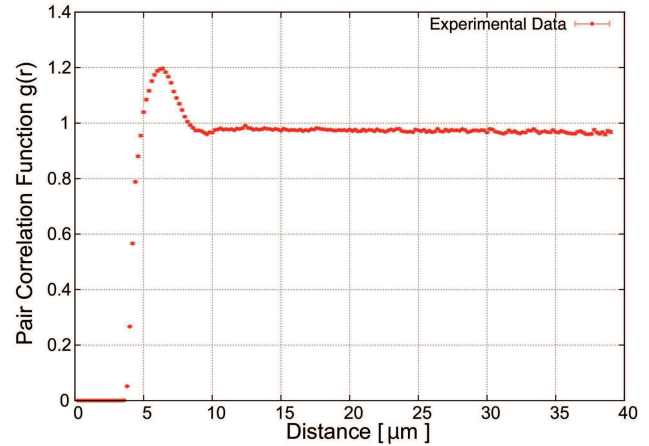}
 \end{center}
  \caption{\label{PairCorr} (Color online) Pair correlation function $g(r)$ has the same form as simple liquids. There is no positional order in the crowd of the Janus particles. Error bars are smaller than symbols.}
\end{figure}

The pair correlation obtained from the experimental data is similar to that of simple liquids (\figref{PairCorr}). It means that there is no obvious ordered structure in the position of the particles like crystals. The profile of the obtained pair correlation is the same as the one observed when particles with finite sizes are randomly distributed.
This profile is also observed in the experiments on the Quincke rollers in the regime called ``polar liquid phase'' where macroscopically directed motion appeared \cite{Bricard2013}. 

\begin{thebibliography}{51}
\bibitem{InterpolationMethod}%
  \BibitemOpen
  \bibinfo {note} {The velocity correlation function is interpolated by pchip method with MATLAB. Different interpolating methods such as spline interpolation and linear interpolation are also used to confirm the robustness of the obtained spectrum. Because the wave numbers that are focused on in \figref{VeloCorr} correspond to length scales larger than the diameter of the Janus particles, the interpolation at the scale smaller than the diameter has little influence on the form of the power spectrum.}
\end{thebibliography}

%

\end{document}